# Classification for the radial component of particles motion in the field of a super-extremely charged object

Valentin D. Gladush, Dmitry A. Kulikov

*Theoretical Physics Department, Dniepropetrovsk National University,*
*Gagarin Ave. 72, Dniepropetrovsk 49010. Ukraine*

**Abstract.** *A method for investigation of neutral test particles motion with the angular momentum L in the Reissner–Nordström field of a super-extremely charged object with mass M and charge Q > M is proposed. The problem of classifying the radial component of motion is reduced to working up a classification for "effective" potential types and subsequent consideration of trajectories for each of these types.*

**Introduction**. The analysis of test particles motion is a classical method for studying the spacetime structure near a gravitating massive object. The electric charge of the gravitational field source, if present, affects substantially the spacetime geometry. Search for particles motion peculiarities that emerge in this case receives considerable attention [1; 6] as it constitutes an important part of the investigation of relativistic configurations.

In the present work we study the radial component of motion for neutral test particles with mass $m$ in the field of a source with mass $M$ and super-extremal charge $Q > M$. This field is described by the Reissner–Nordström metric

$$ds^2 = Fc^2 dt^2 - F^{-1} dr^2 - r^2 d\Omega^2, \tag{1}$$

$$F = 1 - \frac{2\kappa M}{c^2 r} + \frac{\kappa Q^2}{c^4 r^2}, \qquad d\Omega^2 = d\theta^2 + \sin^2\theta\, d\varphi^2. \tag{2}$$

Since the field is static and maintains the spherical symmetry, the total energy of the particle

$$E = c\sqrt{F}\sqrt{m^2 c^2 + F p_r^2 + \frac{1}{r^2}\left(p_\theta^2 + \frac{p_\varphi^2}{\sin^2\theta}\right)} = \text{const} \tag{3}$$

as well as the angular momentum squared

$$L^2 = p_\theta^2 + \frac{p_\varphi^2}{\sin^2\theta} = \text{const}, \tag{4}$$

with $P^\mu = mc\, dx^\mu/ds$ being the particle four-momentum, are conserved. These formulas yield the expression for the radial component of velocity

$$\left(mc^2 \frac{dr}{ds}\right)^2 = E^2 - \left(1 - \frac{2\kappa M}{c^2 r} + \frac{\kappa Q^2}{c^4 r^2}\right)\left(m^2 c^4 + \frac{c^2 L^2}{r^2}\right). \tag{5}$$

**Scale invariance, new parameters and effective potential.** The dynamical system under consideration includes five parameters: $M, Q, m, E, L$. However, the number of essential ones is less than five because the system admits the two-parametric $G^2$ group of scaling transformations

$$E' = \frac{1}{\alpha}E, \quad m' = \frac{1}{\alpha}m, \quad L' = \frac{1}{\gamma}L, \quad Q' = \frac{\alpha}{\lambda}Q, \quad M' = \frac{\alpha}{\lambda}M, \quad r' = \frac{\alpha}{\lambda}r, \quad s' = \frac{\alpha}{\lambda}s. \tag{6}$$

As a new set of parameters and variables we choose the independent invariants of the $G^2$ transformations

$$\varepsilon = \frac{E}{mc^2}, \quad \sigma = \frac{m\sqrt{\kappa}}{c|L|}|Q|, \quad \mu = \frac{m\kappa}{c|L|}M, \quad z = \frac{mc}{|L|}r, \quad \tau = \frac{mc}{|L|}s \tag{7}$$

where $\varepsilon, \sigma, \mu, z, \tau$ are the reduced energy, mass, radius and proper time, respectively. In their terms equation (5) becomes

$$\left(\frac{dz}{d\tau}\right)^2 = \varepsilon^2 - \left(1 - \frac{2\mu}{z} + \frac{\sigma^2}{z^2}\right)\left(1 + \frac{1}{z^2}\right) = \varepsilon^2 - 1 + \frac{2\mu}{z} - \frac{1+\sigma^2}{z^2} + \frac{2\mu}{z^3} - \frac{\sigma^2}{z^4} = -W_V, \tag{8}$$

where $W_V$ is the velocity potential and the condition of super-extremality reads $\sigma > \mu$. The motion regions and turning points can be found from the conditions $W_V \leq 0$ and $W_V = 0$, whereas the circular orbits result from $W_V = 0$, $\partial W_V/\partial z = 0$.

Introducing a suitable potential enables one to facilitate the treatment [3]. For the present problem the energy potential due to Ruffini [4] proved to be convenient. In what follows, it is referred to as the effective potential.

According to [3], the velocity potential (8) gives rise to the effective potential in the form

$$W_\varepsilon^2 = 1 - \frac{2\mu}{z} + \frac{1+\sigma^2}{z^2} - \frac{2\mu}{z^3} + \frac{\sigma^2}{z^4}. \tag{9}$$

Now the motion regions and turning points are determined by the conditions $W_\varepsilon^2 \leq \varepsilon^2$ and $W_\varepsilon^2 = \varepsilon^2$. As a consequence, the classification of motions is connected with distinguishing the different types of the potential $W_\varepsilon^2$.

We shall work out the classification that is governed by the central object parameters $\{M,Q\}$ and the particle angular momentum $L$ translated into parameters $\mu$ and $\sigma$. For this end, let us consider the circular orbits. They can be found from the conditions $W_\varepsilon^2 = 0$ and $\partial W_\varepsilon^2/\partial z = 0$ that amount to algebraic equations

$$(1-\varepsilon^2)z^4 - 2\mu z^3 + (1+\sigma^2)z^2 - 2\mu z + \sigma^2 = 0, \tag{10}$$

$$z^3 - \frac{1+\sigma^2}{\mu}z^2 + 3z - \frac{2\sigma^2}{\mu} = 0. \tag{11}$$

For a stable circular orbit it is necessary to fulfill the minimum condition $\partial^2 W_\varepsilon^2/\partial z^2 > 0$. If there is an inflection point where $\partial^2 W_\varepsilon^2/\partial z^2 = 0$, it corresponds to the last circular orbit.

**Circular orbit radii and their classification.** Let us bring the cubic equation for circular orbits (11) to the standard form

$$z^3 + az^2 + bz + c = 0. \tag{12}$$

$$a = -\frac{1+\sigma^2}{\mu}, \qquad b = 3, \qquad c = -\frac{2\sigma^2}{\mu}. \tag{13}$$

Then the types of its solutions are deduced from the value of the discriminant

$$27D = \left(-a^2/3 + b\right)^3 + \frac{1}{4}\left(2a^3/27 - ab/3 + c\right)^2. \tag{14}$$

Equation (12) may have (i) one real root and a pair of complex conjugate roots ($D>0$); (ii) three real roots from which at least two are equal ($D=0$); (iii) three different real roots ($D<0$). For identifying these cases, we rewrite the discriminant as

$$D = \left(1 - \mu_+^2/\mu^2\right)\left(1 - \mu_-^2/\mu^2\right). \tag{15}$$

$$\mu_\pm^2 = \frac{1}{72}\left(3(1+14\sigma^2+\sigma^4) \pm \sqrt{2}(1-5\sigma^2)\sqrt{(3+\sigma^2)(1-5\sigma^2)}\right). \tag{16}$$

In the plane of parameters $(\mu,\sigma)$, the equality $D=0$ is the boundary of the curvilinear angle $\Sigma^{(2)} = \Sigma_+^{(2)} + \Sigma_-^{(2)}$, which is depicted in Fig. 1 by a heavy line and corresponds to the case of multiple roots ($z_1 = z_2, z_3$ or $z_1, z_2 = z_3$). The point $\Gamma^{(1)} = \Sigma_+^{(2)} \cap \Sigma_-^{(2)}$ where the curves $\Sigma_+^{(2)}$ and $\Sigma_-^{(2)}$ intersect is associated with the triple root ($z_1 = z_2 = z_3$). The part of the region $D^{(3)} = D_1^{(3)} \cup D_2^{(3)}$ that lie outside the angle $\Sigma^{(2)}$ refers to the case of $D<0$, i.e., to three different real roots $z_1, z_2, z_3$. For points in the region $D^{(1)} = D_1^{(1)} \cup D_2^{(1)} \cup D_3^{(1)}$, outside $\Sigma^{(2)}$, we have $D>0$. Here one root is real $z_3$ (a circular orbit) whereas two others are complex $z_1 = m+in$, $z_2 = m-in$. Note that the super-extremality condition $\sigma^2 > \mu^2$ holds underneath the broken line $\sigma^2 = \mu^2$, which corresponds to extremal black holes.

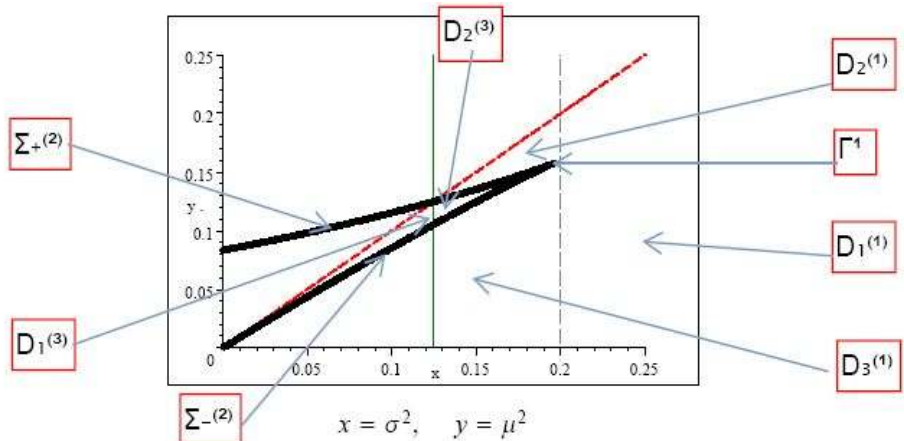

*Fig. 1. Regions in ($\mu$,$\sigma$)-plane that correspond to different types of the circular orbit solutions*

**Parametrization in terms of circular orbits radii and classification of effective potentials.** The derivative of the effective potential (9) can be written as

$$\frac{\partial W_\varepsilon^2}{\partial z} = \frac{2\mu}{z^2}(z^3 + az^2 + bz + c) = \frac{2\mu}{z^2}(z - z_1)(z - z_2)(z - z_3) \quad (17)$$

where $z_1, z_2, z_3$ are the roots of (12). Integrating this equation and taking into account (13), we get the parametrization of the effective potential in terms of the roots of (17)

$$W_\varepsilon^2 = 1 - \frac{2}{2(z_1 + z_2 + z_3) - z_1 z_2 z_3}\left(2 - \frac{z_1 + z_2 + z_3}{z} + \frac{2}{z^2} - \frac{z_1 z_2 z_3}{2z^3}\right)\frac{1}{z}. \quad (18)$$

Note that in case of complex roots one should put $z_1 = m + in$, $z_2 = m - in$. The parameters of the system μ and σ, and the super-extremality condition $\sigma^2 > \mu^2$ take the form

$$\mu = \frac{2}{2(z_1 + z_2 + z_3) - z_1 z_2 z_3}, \quad \sigma = \sqrt{\frac{z_1 z_2 z_3}{2(z_1 + z_2 + z) - z_1 z_2 z_3}}, \quad z_1^2 z_2^2 z_3^2 - 2\left(z_1^2 z_2 z_3 + z_1 z_2^2 z_3 + z_1 z_2 z_3^2\right) + 4 < 0, \quad (19)$$

with $z_1 z_2 + z_1 z_3 + z_2 z_3 = 3$ and $2(z_1 + z_2 + z_3) - z_1 z_2 z_3 > 0$. This representation permits us to reduce the problem of classifying the trajectories to the task of identifying the type of the potential.

The regions of parameters (μ,σ) for the different types of the potential are shown in Fig. 1. Choosing an arbitrary point inside each of these regions and calculating the roots of equation (17), we compose the potential according to (18). The corresponding plots are presented in Fig. 2. Here the allowed regions of motion for particles with given energy ε are the horizontal line segments $W = \varepsilon^2 = \text{const}$ bounded by the intersection points with the curve $W = W_\varepsilon^2(z)$ (turning points).

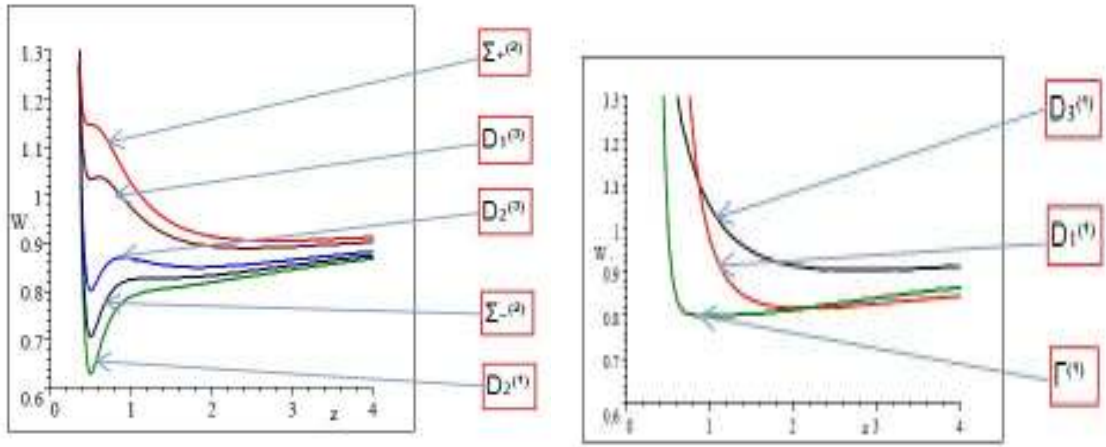

*Fig. 2. Effective potential vs. radius z for different choices of parameters (μ,σ)*

In Table 1 we list the values of circular orbit radii $z$ and particle energies ε calculated with parameters (μ,σ) selected from the regions $D_1^{(3)}, D_2^{(3)}, \Sigma_-^{(2)}, \Sigma_+^{(2)}, \Gamma^{(1)}, D_1^{(1)}, D_2^{(1)}, D_3^{(1)}$ corresponding to different signs of the discriminant $D$ for the case of the super-extremely charged object $\sigma^2 > \mu^2$.

*Table 1. Circular orbit radii and energies of particles for selected values of parameters*

|   | D>0 | | D=0 | | | D<0 | | |
|---|---|---|---|---|---|---|---|---|
|   | $D_1^{(3)}$ | $D_2^{(3)}$ | $\Sigma_-^{(2)}$ | $\Sigma_+^{(2)}$ | $\Gamma^{(1)}$ | $D_1^{(1)}$ | $D_2^{(1)}$ | $D_3^{(1)}$ |
| μ | 0.314 | 0.356 | 0.373 | 0.294 | 0.40 | 0.447 | 0.387 | 0.3 |
| σ | 0.334 | 0.382 | 0.398 | 0.318 | 0.447 | 0.632 | 0.410 | 0.354 |
| z | 0.5, 0.6, 2.455 | 0.5, 0.9, 1.821 | z1=0.5, z2,3=1.3 | z1,2=0.5, z3=2.75 | z1,2,3=1 | 2.111 | 0.5 | 2.778 |
| ε | 1.017, 1.019, 0.944 | 0.896, 0.933, 0.538 | 0.84, 0.91 | 1.071, 0.954 | 0.894 | 0.903 | 0.793 | 0.951 |

**Conclusion**. The proposed method combines analytical and numerical calculations and enables us to work out the characteristics of test particles motion in the field of the charged object. The peculiarity of this motion is that there may exist bound states with energy greater than the rest value ($\varepsilon > 1$). For example, this occurs for the effective potential from $D_3^{(1)}$ (see the left part of Fig. 2) where the finite particles motions for the object with $\mu = 0.314$, $\sigma = 0.334$ are allowed, so that the particles with energy $\varepsilon = 1.017$ move along the circular orbit with radius $z = 0.5$. As seen from Fig. 2, the curve $\Sigma_+^{(2)}$ for the object with $\mu = 0.294$, $\sigma = 0.318$ has an inflection point at the radius $z = 0.5$ that corresponds to the last stable circular object with energy $\varepsilon = 1.071$.

This work was supported by the grant of the "Cosmomicrophysics" program of the Physics and Astronomy Division of the National Academy of Sciences of Ukraine.